# Awareness of self-control

Mohammad Mehdi Mousavi[a], Mahdi Kohan Sefidi[b], Shirin Allahyarkhani[c]

Faculty of Economics, Khatam University, Tehran, Iran[d]
[a,d]m.mosavi@khatam.ac.ir,  [b,d]m.kohansefidi@khatam.ac.ir,
[c,d]sh.allahyarkhani@khatam.ac.ir

February 2023

**Abstract.** Economists modeled self-control problems in decisions of people with the time-inconsistence preferences model. They argued that the source of self-control problems could be uncertainty and temptation. This paper uses an experimental test offered to individuals instantaneous reward and future rewards to measure awareness of self-control problems in a tempting condition and also measure the effect of commitment and flexibility cost on their welfare. The quasi-hyperbolic discounting model with time discount factor and present bias at the same time was used for making a model for measuring awareness and choice reversal conditions. The test showed 66% awareness of self-control (partially naïve behaviors) in individuals. The welfare implication for individuals increased with commitment and flexibility costs. The result can be useful in marketing and policy-making fields design precisely offers for customers and society.

## 1    Introduction

Consider a decision-maker who prefers a larger-later than a smaller-sooner reward. When time passes on and the smaller-sooner reward becomes immediate, the decision-maker reverses their choice and picks the immediate reward. For example, consider a student has decided to participate in a pre-exam session class, in the morning at 8 o'clock, but when the morning arrives, the student prefers to stay in bed and enjoy more time sleeping. These actions are known to be self-control problem. Laboratory experiments show choice reversal behavior in humans and animals. Read and Leeuwen (1998) asked individuals if they preferred fruit or chocolate to have for the next week. Although 74% chose fruit at the time they were asked, in the future, they reversed their choices, and 70% chose chocolate. Also, there is evidence of choice reversal and present bias in economics and financial issues. Intertemporal preferences with these features show self-control problems and diminishing impatience. When decision-makers evaluate outcomes in the distant future, such as quitting smoking, they plan to act patiently. However, when the future becomes close, they act impatiently. Diminishing impatience behavior has been discussed in the literature by two approaches. On one side of the argument, some studies explain diminishing impatience (present bias effect) by introducing an uncertain exponential discounting model (Sozou 1998, Azfar 1999, Dasgupta and Maskin 2005, Halevy 2008). In those studies, the origin of choice reversal is the risk that any future reward contains. On the other hand, some previous works indicate that the presence of diminishing impatience behavior is due to the temptation generated by dynamically inconsistent preferences. As a result, individual awareness of diminishing impatience can lead to demand for a commitment device (Strotz 1973, Akerlof 1991, Elster 1979, O'Donoghue and Rabin 1999a). The results of some experimental studies show the effect of temptation on preferences for commitment (Gul and Pesendorfer 2001) and the impacts of commitment on improving individuals' performance and reducing potential costs of self-control (Toussaert 2018) although both sides of the argument agree on the shape of the discounting function (hyperbolic and quasi-hyperbolic discounting functions), the origin of such behavior is quite different in each of the approaches. One way to distinguish between the origins of choice reversal behavior is by looking at the decision-makers' demand for flexibility and commitment devices. If the origin of diminishing impatience is the only risk (and not temptation), individuals should be unwilling to commit. There is also experimental evidence that shows the existence of demand for commitment (Casari 2009). However, the effects of uncertainty and new information on choice reversal cannot be ignored (Casari and Dragone 2015). In this study, we tried to measure the decision maker's awareness of choice reversal behavior by analyzing demand for commitment and flexibility, according to the quasi-hyperbolic discounting model through an experiment. it will be used temptation and uncertainty as behavioral factors in quasi-hyperbolic model. Also, in this experimental study what has been evaluated is the effects of commitment and flexibility on welfare implications. We used some sweets rewards for small and sooner and foods for large and later rewards. Sweet is a good option to tempt people until appearing their present bias in choice. Also, money was taken for flexibility and commitment demanders. We measure the utility of people with a quasi-hyperbolic model in a short time with a large present bias and compare that with expectation utility in a long time. These comparisons help us first to measure impatience then predict choice reversing in individuals and determine a person's type in awareness of self-control problems. Also, the demand for commitment devices (money in our test) will help us to answer whether commitment will increase the welfare of people or not. The result of research help policymakers and also economist design model for increasing and adjusting the preferences of society. We also mention some limitations in our research and test that affect our results like flexibility welfare and temptation value. These are so good subjects for further research. We run the model with our test results and then compare it with another test of demand for flexibility and commitment that measure with rewards just money (Casari 2009) and report the result of it.

The paper is structured as follows. Section 2 previous literature explain self-Control problem resources and basic models that use in this experimental model, Section 3 explain the model of research, Section 4 is about experiment and execution model of it, Section 5 is the result of experiment, Section 6 is the result of the research and section 7 is conclusion.

## 2      Previous literature

The main model provided is a continuation of the existing models in the subject research self-control problem, temptation, and self-awareness in this section, each of these subjects and the model presented by them will be examined, and in the next sections (the model section) the model will be developed and presented in the continuation of the past research.

### 2.1     Self-control problem and temptation

It's possible to have been several sources for self-control problem one of them is diminishing impatient. As an example, if a person faces two types of choice, the first type is the selection between studying last night before the exam or going to a party. the second type is the decision about going to a party or studying exam last night before next week exam, select party in the first type and studding in second type exhibit diminishing impatient (Halevy 2008). Another source of self-control problem is temptation, for example, a person decides to eat some healthy things, when she is not hungry, like fruit and vegetables but when she is hungry and faces fast food, she reverses her choice (Gul and Pesendorfer 2001). We can say procrastination is also source of self-control problem, in a story consider a person has to send a request for a loan and delay in sending the request has just cost for that person but he doesn't do that as scheduled and postpone it, i.e., individuals $LLC \succ SSC$ prefer large later cost to small sooner cost choice. These are some source effects on our decision with self-control problem (Akerlof 1991). Gul and Pesendorfer (2001) argue that people have "temptation preferences", their model consider that intertemporal choices in any framework are related to a person's welfare (Gul and Pesendorfer 2001). They conclude that people who are tempted in decision time and also avoid tempter option, they select the better choice (Large Later). Also, they select pre-commitment to avoid them from temptation in time of decision making. It shows that some people are aware of their temptation in their decision-making process. Temptation in decision-making maximizes instantaneous utility and precommitment can improve total utility during decision making. The demand for pre-commitment needs some degree of self-awareness that was discussed in the previous section. For relating temptation to self-control, we couldn't use time-consistence and we should use the dynamic discounting model with quasi hyperbolic formulate (discussed more in detail further on). Strotz (1955) and Pollak (1968) have studied consumption choice under temptation by (Laibson 1997, Strotz 1973, Pollak 1968). Quasi hyperbolic agents have a preference for commitment if they exhibit choice reversal (Lipman and Pesendorfer 2013). Subjects face temptation in their selection but when is offered commit for resist tempting most subjects select it the first but some of them select with delay (Houser, et al. 2018).

## 2.2 Self-awareness

People may not be aware that, their preferences may change over time if they have time-inconsistent preferences. Strotz (1955) and Pollak (1968) argue about an individual's awareness in extreme forms. Some individuals can completely predict their decisions in the future from now, while other individuals can predict how their decisions may change over time. The first group mentioned above is called, "*naïve*" and the second group of people mentioned is called "sophisticated". While researches show that people are between two extreme states, people are not completely "*naïve*" or completely "sophisticated", they can be "*partially naïve*". For identifying sophistication in individuals, it should be attended on demand of commitment in them. 'Sophisticated' people expect to change their decision over time and then they usually demand commitment. They know that temptation will be affecting their decision in the future. They avoid the temptation of changing decisions with commitment. For example, an overweight individual may decide to eat a vegetable lunch. Because she knows that may change her decision in the time of ordering food, please from one of her friends that she order food instead of herself and just order a vegetable food (never order fast food or fat even if herself changed her decision). This is like a commitment to avoiding temptation in the future. Several kinds of research have been done about the demand for commitment in individuals. This examination shows that sophisticated people tend to have more complex behaviors and what's the relation of overconsumption and addiction with the degree of sophistication and degree of awareness. O'Donoghue and Rabin (2001) present a model between extreme states of individuals awareness which is not naïve or sophisticate, and this is called *partially naïve* (O'Donoghue and Rabin 2001). *Partially naïve* individuals are defined as a person who is aware of their self-control problems, but they are not aware of the magnitude of their self-control problems. Procrastination source is a partially naïve person who is not completely sophisticated. Policymaking is very effective in the life of the average citizen and the welfare of their country. Making good policies requires there to be vast amount of knowledge on the decisions of the citizens and their behaviors in different conditions. The different categories of citizens and the measuring of sophistication and naïveté can help in the decision process of policy making and policy implementation.

## 2.3 Self-awareness and dynamic inconsistence choice

Economists' models before O'Donoghue and Rabin's (2001) model assumes that the intertemporal preference is time-consistence (O'Donoghue and Rabin 2001). O'Donoghue and Rabin argue that intertemporal preferences take on the specific form of time inconsistency, individuals have self-control problems caused by a tendency for immediate gratification and this model separates people into three different groups with respect to their knowledge about their self-control problems. Naïve people are not aware of their self-control problems, while sophisticated people are fully aware of their self-control. A model that is more realistic than the extreme state of naivete and sophistication is *partially naïve*. *Partially naïve* people know that they have self-control problems in the future but they are not aware of the severity of their problems. The individuals' expectation of self-control parameters that O'Donoghue and Rabin modeled, is given by:

$$\widehat{U}_t(u_t, u_{t+1}, u_{t+2}, \ldots, u_T) \equiv \delta^t u_t + \hat{\beta} \sum_{\tau=t+1}^{T} \delta^T u_\tau \qquad (9)$$

In this function, $\widehat{U}_T$ is the expectations of total utility that the decision-maker estimate, and $\hat{\beta}$ denotes the decision-maker estimation of $\beta$. In other words, by considering $\hat{p}$ as the decision-maker estimation of the probability that she will reverse her choice, can be represented $\hat{\beta}$ as $\hat{\beta} = \hat{p}\beta$.

Hence, individuals are divided into three categories according to their $\hat{\beta}$ as follows:

**Fully naïve**: If an individual is a fully naive decision-maker, her expected utility function does not contain the self-control parameter ($\hat{\beta} = 1$).

**Sophisticated**: If an individual is a sophisticated decision-maker, her estimated self-control parameter is exactly equal to the real self-control parameter ($\hat{\beta} = \beta$).

**Partially naive**: If an individual is partially naive, her estimated self-control parameter is lower than the real self-control parameter ($\beta < \hat{\beta} < 1$).

In this research, we use previous concepts to design a model for measuring awareness with respect to self-control problems. In our model, we used a dynamic inconsistent model. For the test self-control problem as we said there are several agents, one of them is the temptation that is used in this model for making present bias in the participants. The type of participants in awareness with respect to this type of bias is considered partially naïve. All decisions have some uncertainty in the result then people never can say exactly what will happen in the future.

we want to design a model for measuring the probability that is hidden in our mind with respect to our behavior in the future and how much is near to our decision. In the previous models, there is no method that measures awareness of self-control problems.

# 3 Model

To explain the model, consider a decision-maker who wants to choose between two options A and B, where A involves becoming a smaller-sooner reward (SS) and B includes becoming a larger-later reward (LL)[1]. Suppose that the individual chooses B (LL reward) in the decision time ($t_0$) but when the near future comes ($t_1$), she changes her mind and prefers option A (SS reward). This action is known as "*choice reversal*" or self-control problem. Now the question is, how much does the decision-maker know about this problem? and if so, does the individual have a demand for a commitment device? To answer these questions, the model is presented in three parts.

## 3.1 Intertemporal preferences

The decision-maker intertemporal preferences by dividing the time into the decision time ($t_0$) and the time to receive the SS reward ($t_1$) and the LL reward ($t_2$), based on (utility function) equation (6) is defined by:

$$U = u(C_0) + \beta\delta u(C_1) + \delta^2\beta u(C_2), \ 0 < \delta < 1, \ 0 < \beta < 1 \quad (10)$$

where $u_t$ in $t = \{t_0, t_1, t_2\}$ the period is utilities discounted by $\beta\delta^t$ factor that $\delta$ is her discount factor and $\beta$ will denote the present bias parameter. Note that if $\beta = 1$, the effect of inconsistency is ignored and the model only reflects the time-consistent preferences. By substituting the utilities of choosing options A at $t$ with utility $U_A^{(t)}$ and B at $t$ with utility $U_B^{(t)}$ into the equation (10), it's readily obtained that the agent's intertemporal preferences at $t_0$ is equal: (Proof 1 at appendix 1)

$$U_A^{(0)} = \beta\delta U_{SS}, \ U_B^{(0)} = \beta\delta^2 U_{LL} \quad (11)$$

and similarly, at $t_1$ are equal:

$$U_A^{(1)} = U_{SS}, \ U_B^{(1)} = \beta\delta U_{LL} \quad (12)$$

where $U_{SS}$ represents the utility that a decision-maker achieves when receiving the SS reward and $U_{LL}$ represents the utility of receiving LL reward (assuming that $\beta$, $U_{SS}$ and $U_{LL}$ are exogenous variables). Consequently, it is easy to show from (11) and (12) that the agent prefers B at $t_0$, if and only if $U_B^{(0)} > U_A^{(0)} \Leftrightarrow U_{LL} > \frac{U_{SS}}{\delta}$ and prefer A at $t_1$ if and only if $U_A^{(1)} > U_B^{(1)} \Leftrightarrow U_{LL} < \frac{U_{SS}}{\delta\beta}$. Hence choice reversal can be occurred in this example only if $\frac{U_{SS}}{\delta} < U_{LL} < \frac{U_{SS}}{\delta\beta}$. (Proof 2 at appendix 1)

---

[1] LL is a large value reward with some delay and SS is a small value reward without delay

## 3.2 Expected utility

As discovered earlier, the agent's preferences in each period and the necessary condition for having the choice reversal were obtained. Now the question is, to what extent is the decision-maker aware at $t_0$ that she might be tempted in condition at $t_1$ and reverse her choice[1]. In the line with this, let $E\left(U_{B\,\text{flex}}^{(0)}\right)$ represent the expected utility of choosing B at $t_0$ with flexibility option:

$$E\left(U_{B\,\text{flex}}^{(0)}\right) = \hat{p} U_A^{(0)} + (1-\hat{p}) U_B^{(0)} \qquad (13)$$

where $\hat{p}$ represents the probability that the agent expects to be tempted at $t_1$ and reverses her choice from B to A and consequently $(1-\hat{p})$ denotes the expected probability of not to be tempted at $t_1$ and stick to her choice. The expected utility of individuals based on the type of category in which they are placed is for a sophisticated person $\hat{p} = 1$, A fully naïve person $\hat{p} = 0$, And for *partially naïve* person $0 < \hat{p} < 1$. We named $\hat{p}$ awareness of self-control problem in this paper.

## 3.3 Commitment device and flexibility

Now we want to evaluate the demand of people who are aware of this problem to have a commitment device[2] to this end, suppose that in the previous example two preconditions are represented to the decision-maker at $t_0$, the first precondition is that the agent be able to be flexible in her choice at $t_1$ and can reverse her opinion regardless of her choice at $t_0$ and the second one is that she constraints her choice by having a commitment device and cannot change her mind at $t_1$ the purpose of this is to help the agent to resist her temptation at $t_1$ by commitment device. With this in mind that the selection of commitment device will naturally have an implicit cost which is losing the flexibility option $(V_f)$[3]. Therefore, the expected utility of choosing B at $t_0$ when the agent uses commitment device is obtained:

$$E\left(U_{B\,c}^{(0)}\right) = \beta \delta^2 U_{LL} - V_f \qquad (14)$$

and if the commitment device is costly, this will satisfy:

$$E\left(U_{B\,c}^{(0)}\right) - M = E\left(U_{B\,\text{flex}}^{(0)}\right) \qquad (15)$$

such that M is the maximum utility unit that the decision-maker is willing to pay for having a commitment device at $t_0$. In other words, paying M units of utility at $t_0$, makes the agent indifferent between playing the game with or without the commitment. Likewise, a condition where the flexibility is costly can be considered:

$$E\left(U_B^{(0)}{}_{\text{flex}}\right) - N = E\left(U_B^{(0)}{}_c\right) \tag{16}$$

that N implies the maximum utility unit that the decision-maker is willing to pay for having a flexibility at $t_0$.

---

[2] In this study, the behavior of partially naive persons is examined
[3] Casari (2009) argued that commitment deprives the agent of the possibility to change her opinion under the new information and uncertainty.

### 3.4 Measuring the Awareness of self-control problem

In order to estimate the awareness of self-control problem ($\hat{p}$), these relations are obtained with respect to the equation (15):

$$\hat{p} = \frac{M + V_f}{\delta^2 \beta U_{LL} - \delta \beta U_{SS}} \tag{17}$$

with respect to the equation (16):

$$\hat{p} = \frac{N + V_f}{\delta^2 \beta U_{LL} - \delta \beta U_{SS}} \tag{18}$$

See (Proof 3 at Appendix1)

Note that variables $M$, $N$, $V_f$ and $\beta$ are assumed to be exogenous and in our model, we consider $V_f = 0$ because of complexity in measuring but actually it's not zero[4]. It is clear that according to the equations (10) and (11), variables $M$ and $N$ are maximized if and only if $\hat{p} = 1$ and hence their values are equal to $\delta^2 \beta U_{LL} - \delta \beta U_{SS}$. If a decision maker is ready to pay more, she overestimates the self-control parameter. Consequently, with having the parameters $M$ and $N$ of each person (Collected from experiment results), the awareness of the self-control problem parameter can be easily calculated.

### 3.5 Welfare implications

Given all the above, the effect of proposing a commitment device to the decision-maker who choose option B at $t_0$ on her welfare can also be examined as follows:

$$WI = E\left(U_B^{(0)}{}_c\right) - E\left(U_B^{(0)}{}_{\text{flex}}\right) \tag{19}$$

The welfare implication of individuals in this model is equal to M, which means that the welfare of people as same as commitment device will increase in their choice.

---

[4] It's a new subject for future research, measuring the value of implicitly welfare loss.

## 4  Numerical approach of the model

The model can be implemented in three steps. We use a numerical approach to describe each part of the model. The experimental test that will explain in the following sections has used food and sweet price as reward.

**Step1:** Each decision has four parameters: The amount of the two possible payments and their delays. Both payment amounts were held constant throughout the procedure at $100 and $110. By using the data from the first part and modelling it with the quasi-hyperbolic discounting model, we get (labelling U as the value of each option and assuming a risk-neutral utility function)

$$SS = (100, \epsilon) \Rightarrow U_{SS} = 100$$
$$LL = (110, D^*) \Rightarrow U_{LL} = 100.\beta.\delta^{D^*}$$

$D^*$ has been increased to temp the decision maker choose $U_{SS}$ in this point maximum of $D^*$ give minimum of discounting rate $(1 + \delta_{max})$.

$$U_{SS} > U_{LL} \Leftrightarrow 110.\beta.\delta^{D^*} < 100$$

$$\delta_{max} = \sqrt[D^*]{\frac{SS}{LL.\beta}}$$

**Step2:** Both rewards are added with a front-end delay (FD$^*$) in order to detect possible choice reversals. The smaller-sooner reward maintains a consistent delay of FD$^*$+$\epsilon$ days and the larger-later reward maintains a delay of FD*+D* days, which ensured throughout part two that the waiting time difference between SS and LL was constant. By modelling part two based on quasi-hyperbolic discounting, we get

$$SS = (100, FD^* + \epsilon) \Rightarrow U_{SS} = 100.\beta.\delta^{FD^*}$$
$$LL = (110, FD^* + D^*) \Rightarrow U_{LL} = 110.\beta.\delta^{(FD^* + D^*)}$$

$$U_{LL} > U_{SS} \Rightarrow 100.\beta.\delta^{(FD^* + D^*)} > 100.\beta.\delta^{(FD^* + \epsilon)}$$

From the inequalities from steps one and two, we get

a) $100 > 110.\beta.\delta^{D^*}$

b) $100.\beta.\delta^{FD^*} < 110.\beta.\delta^{FD^*+D^*}$

$$\Rightarrow \frac{100}{110} < \beta < 1, \sqrt[D^*]{\frac{100}{110}} < \delta < 1$$

$$U_{LL} = \beta.\delta_{max}^{(FD^*+D^*)}.LL$$

$$U_{SS} = \beta.\delta_{max}^{FD^*}.SS$$

Up to this point, the discounting elements $(\beta, \delta)$ for quasi-hyperbolic and hyperbolic discounting are measured.

**Step3:** This step includes decisions aimed at measuring the preferences for commitment and flexibility. More precisely, Subjects should choose between two options. Each question in part three aims to detects demand for commitment, demand for costly commitment or demand for flexibility.

**Step 3-1:** We have two options:

**Option A:** $SS = (100, FD^* + \epsilon)$ or $LL = (110, FD^* + D^*)$

**Option B:** $LL = (110, FD^* + D^*)$

If a decision maker chooses option B, she sacrificed the utility of having flexibility for a commitment device. By labelling the value of the flexibility option as $V_f$, the values of each option are equal:

$$U_A = 100.\hat{p}.\beta.\delta^{FD^*} + 110.(1-\hat{p}).\beta.\delta^{FD^*+D^*}$$
$$U_B = 110.\beta.\delta^{FD^*+D^*} - V_f$$

If the individual prefers option B to A: $U_A > U_B \Rightarrow \hat{p} > \frac{V_f}{U_{LL}-U_{SS}}$. If the individual prefers options A to B: $U_A < U_B \Rightarrow \hat{p} < \frac{V_f}{U_{LL}-U_{SS}}$. When we find indifference point $(\delta_{max})$ then: $\hat{p} = \frac{V_f}{U_{LL}-U_{SS}}$.

**Step 3-2:** If a decision maker chooses option B, she sacrificed the utility of having flexibility and pays M for a commitment device. By labelling the value of the flexibility option as $V_f$, the values of each option are equal: $\hat{p} = \frac{M+V_f}{U_{LL}-U_{SS}}$.

**Step 3-3:** If a decision maker chooses option A, she pays N for flexibility. By labelling the value of the flexibility option as $V_f$, the values of each option are equal: $\hat{p} = \frac{N+V_f}{U_{LL}-U_{SS}}$.

# 5  Experiment

In the behavioral sciences, experiments should minimize behavioral biases and focus on specific behavior. Such experiments have complexities, so we should near the appropriate design step-by-step. In this field, we are faced with different individuals with different experiences, characters, and personalities. The design of an experiment must be able to consider all aspects of individuals and their commonalities and also avoid the influence of external factors on the results as much as possible. In this research, we understood that several experiments should have been done until we achieve a good method. Experimenting with several different ways and methods help us to minimize its behavioral biases. One of the interesting experiences gained in this direction was that some previous researchers believed that behavioral experiments could be hypothetically tested and the results of hypothetical testing would be the same as actual testing, but in this research, we conducted that the opposite was evident. In the following sections, some of them will be explained.

## 5.1  Experiment features

The experiment should measure the amount of awareness of self-control and its effect on a person's behavior, and therefore there should be a temptation in the test prize. The temptation of human beings can take many forms. This experiment has used one of the basic needs of people according to the category of needs Maslow the physiological needs of people include air, food, drinking, shelter, clothing, heat, sleep, and sexual needs (Gawel 1997). In this research, the temptation to enjoy food is used. This type of temptation has a lot of power and people need some self-control to avoid it. The target population was engineering and economics students in the age range of 20 to 39 years. This type of award operates almost independently of the gender and personality of the individual. Another point that should be taken into account is that people may have different food tastes and therefore the degree of utility is also different. For this case, conditions were considered in the test that is independent of this variable, a part of the test through Sweets were considered (preferred by the majority of people) and for the other part of this test, the food shopping prize is awarded from an online food preparation platform and the price of each food is qualitatively expressed to people that it has almost the same utility among all participants. Another condition that should be considered in the test is the equilibrium point based on the game theory that the individuals face when making a decision (Colman 2003). Some researchers argue that presenting the actual amount of the award is not necessarily in the test results, our experiments show that the actual amount of the award in this test can create a real sense in the people and this feeling reveals the more precise desire of the people being tested (reduce bias).

**5.2    Experiment design**

For reaching the final opinion for choosing the test method, various selection processes were carried out. We test three methods for measuring test online game, picture method and the last one real reward.

First test, it was designed for the computer game test and the test was done through this game. Story of game was about a farmer that sold raw products at the moment or processed in the future (it's like SS and LL in model). In the process of the game, we had some bias that affect on the result of our test one of the challenges we encountered while running the test through this game was that the game reward did not tempt individuals, and people tried to maximize their reward regardless of the delay. other biases were *anchoring bias* and *framing effect* that make error in our measurement. Numbers in the game effect on subjects choose. Shape of products can't make real sense at individuals Therefore, despite these challenges, the use of the online game in the test was failed.

The next method used to perform the test was a set of tempting images of foods and people eating with questions. In this method, people first see the images, and then the questions are asked through the test taker. In this method was failed, due to the lack of necessary temptation in people through experiment's images and attention of people to other elements of picture.

According to the experience gained from previous methods and the behavior of individuals, such as the experimental test was taken in Casari's test (Casari 2009), It was agreed that the test should be implemented on individuals in practice. The practical test was performed in such a way that rewards items were prepared for the test subjects and given to them with their choice. The challenges we faced in previous tests were no longer observed in this method and the amount of temptation that was necessary was made in people. We prepared an environment for each person that was a delicious sweet in front of them and questioner was there in front of them. About the questions, the next section will explain more. In the design of questions based on the model section, we need to access the minimum value of the discount rate then we tried with increasing delay in the test to find the indifference point (minimum value of discount rate) for example when the subject says indifference point is 5, we ask a question about 6 days and so on until stop in fix delay and they don't increase it. The test has two rewards one of them is cheaper and sooner (sweet that is the same between subjects but maybe different as type with their taste) and another has a high price (credit of food) and later. The second stage of the test is designed for recording the behavior of individuals with a small effect of present bias. All the stages of the test are in one session and about 15 minutes.

Another parameter that was used in the test was commitment and flexibility options. These parameters were also questioned at each stage. Demand for commitment can show awareness of the subject with respect to "choice reversal" action or high awareness. Cost of commitment and flexibility decreased credit of their reward (LL) in the future according to model.

### 5.3  Experiment execution

In the first stage according to figure 1, the food credit and sweets options preferred by each subject were prepared, then subjects had to choose between the available sweet items and future food credit with a delay of D days. If the person chooses sweets, the test will finish for that person. Choosing sweets at the beginning indicates the temptation in the person. If a person chooses food after D days, the amount of variable D increases with 1-day steps to touch the point that the person reverses her choice and choose sweets. If a person chooses food after D days, the amount of variable D increases to touch the point that the person reverses her choice and eat sweets. This number of days is indicated by D*, which shows the point of indifference between the choice of food and sweet[5]. After selection and determining D* for each person, the examiner will ask each person: "If sweet is served in presence of you by others may tempt you and you change your decision before D* days (when you change your decision, food credit will be canceled) then you eat sweets, and loss food". Now two options are proposed for the subject, first one is: *"Does the examiner prevent you from being tempted?"* If the answer was yes, the next question is *"whether he is willing to pay the examiner for this prevention?"* if his answer was negative (No), the second question is *"whether you will want freedom of choice i.e., choice reversal before D*?"* And if the answer was yes, another question raised was *"whether you are willing to pay for the freedom of choice?"* Then, after recording these comments, the second stage of the experiment will be held.

**Figure 1.** First stage of experiment

---

[5] D* is special for individuals according to the first stage of the experiment

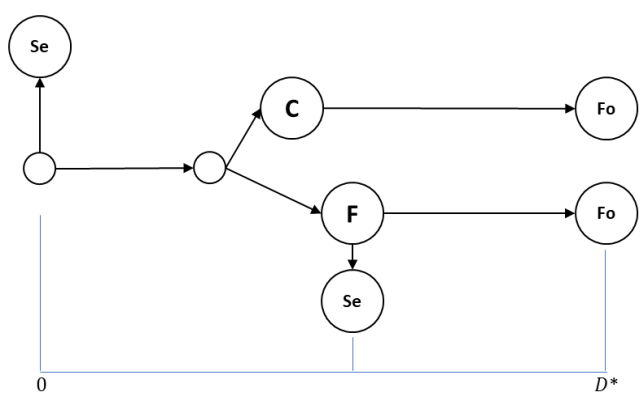

[Se] show Sweets, [Fo] show Food credit, [C] show commitment and [F] show flexibility in options. Also, time line is from 0 day to D* days.

In the second stage of the experiment, the question was asked for the subject to choose between sweets after D Days and food credit after D + D* days that we named this variable forward delay (FD*). The purpose of this step is to examine the changes in temptation created in the individual based on the hyperbolic discounting model. The selection of people will be evaluated according to the choices of the first stage, the second stage will be held only for those who have chosen food in the first stage, and if people choose sweets in the second stage, we will increase the number of D* days to choose the food. The value of D* that the subject changed her choice will be selected as the indifference of that person in the future. The payment method that was used in the first stage for the selection between sweets and food was real.

**Figure 2.** First stage of experiment

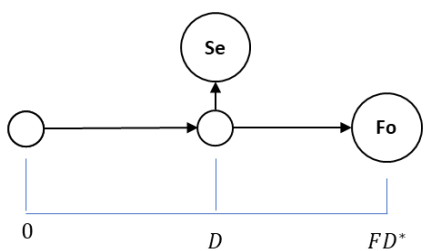

In the first stage sweet credit paid immediately (type of sweet was a Turkish sweet that is popular in Iran called "Baklava" which is about four or six slices in a pack) and food credit sent in an online platform of food ordering "snapp food" [6] in Iran at D* indifference time. In the second stage we paid just sweet for the people but for food in the second stage payment method was hypothetical and just paid for 28% (37 individuals). We use some temptation method The temptation method for individuals with delays was messaging individuals on social media with a picture of sweets at different times for individuals that didn't want commitment (costly or costless) and people that want to change their decision answered us in that.

---

[6] Online platform food ordering https://snappfood.ir/

### 5.4    Experiment of Casari (2009)

Casari (2009) has studied the time preferences of 120 subjects recruited from the undergraduate population of Jaume I University of Castellon, Spain. Each subject faced a series of choices between a smaller-sooner payment (SS) and a larger-later (LL) payment with delays between 2 days and 22 months. Choices were divided into three parts: a first part to measure impatience, a second part to detect choice reversal, and a third part to assess preferences for commitment and flexibility. In Casari's test reward is pure money and it's so interesting for us, then we map our model to his data and analysis the result of that. This research just used data from the subject's that changed decisions along test execution. Also, we considered the observation of his test as subjects. Approximately 82 persons are in the group of people that reversed their choice in his test.

The test of Casari has four parts that we used three parts of his test. In part one, each decision has four parameters: The amount of the two possible payments and their delays. Both payment amounts were held constant throughout the procedure at $100 and $110. The goal of part one was to elicit an approximate measure of impatience $D^*$ by fixing the delay of the smaller-sooner payment at 2 days, $SS = (\$100, 2\ days)$, and varying the delay D of the larger-later payment, $LL = (\$110, D)$, up and down until the subject was approaching indifference between the two payments. Part two offered a commitment to subjects for taking option B (LL option with $110 reward) and also give some costly commitment options (cost of commitment was $2). In part three Casari tests individuals with a flexibility option (the cost of the flexibility option was also $2). For mapping data of Casari's test, we used data from Table1 part of Strict commitment and Flexibility, and Table 6 part of Willingness per day. Table 6 gives information about the D parameter in our test then we used 0-4 days as SS selection and more than it as delay for indifference point. We convert the price of the dollar to our test environment price (Rials) with a convert rate of 20000 Rials for each USD unit. Table1 of his paper give us the result of individual's choice in various types like costly commitment, costless commitment, and flexibility (Casari 2009).

### 6    Results

In this section, we review the results of the awareness of self-control that has been tested and then we model the result of Casari's (2009) test data and for individuals that reversed their choice in his test. In this experiment, 136 participants have been tested. The experiment was conducted among students of SUT [7], AUT [8], SRB [9], and Khatam University [10] in the age range of 20 to 35 years. The test results consist of two parts. The first part is the statistical results of the data collected from the test and the second part is the result of the calculations performed and also the relationship between the test variables. According to Table 1, 59% of people are male and 41% are female. Among the test subjects, 8.82% chose SS, 18.38% chose LL with costly commitment, 8.09% chose LL with costless commitment and 64.71% chose LL with flexibility.

**Table 1.** Subjects respect to selection reward types.

| %Gender (N) | Reward type | %Subject (N) |
|---|---|---|
| 41.17% Female (56) | SS | 8.82% (12) |
| | LL with costly commitment | 18.38% (25) |
| 58.82% Male (80) | LL with costless commitment | 8.09% (11) |
| | LL with flexibility | 64.71% (88) |

According to Table 2 of the results obtained from the first stage of the test, the values of D* individuals, which indicate the number of days those individuals are indifferent to receiving SS and LL, have an average of 11.72 for all testers with std. 16.03. And FD* values people have an average of 11.23 and deviation from the criterion of 10.38. Also, the average cost of commitments is 4640, and the average cost of flexibility is 4265.3. The count of people that don't pay any cost is 48 people.

**Table 2.** Results obtained from the first and second stage of the test

| Variable | Min value | Max value | Average | Std. |
|---|---|---|---|---|

---

[7] Sharif University of Technology, website: https://en.sharif.edu
[8] Amirkabir University of Technology, website: https://aut.ac.ir
[9] Science and Research Branch, website: https://srb.iau.ir
[10] Khatam University, website: https://khatam.ac.ir

|  | 1 | 90 | 11.72 | 16.03 |
|---|---|---|---|---|
| D* | 1 | 90 | 11.72 | 16.03 |
| FD* | 1 | 50 | 11.23 | 10.38 |
| Commitment's cost | 1000 | 10000 | 4640.0 | 3094.08 |
| Flexibility's cost | 1000 | 10000 | 4265.3 | 3060.33 |

According to Table 3, among those who had self-control and were aware of it, 36.48% of subjects have less than 50% aware and 31.08% subjects have more than 50% awareness, so a large percentage of people showed *partially naive* behavior with high awareness. Individuals whose $\hat{p}$ is between 0 and 0.5 show low awareness and observed 24 people with small value of $\hat{p}$ that reversed their choice along test execution. According to a study conducted in (Imai, et al. 2021) research, the amount of PB[11] in individuals is about 0.8 to 0.95 and preferred 0.88 in average, and therefore the values of $\hat{p}$ in terms of that calculated.

**Table 3.** Results obtained from the first and second stage of the test

| $\beta = 0.88$ | |
|---|---|
| Range $\hat{p}$ | %Subjects (N) |
| $0 < \hat{p} < 0.5$ | 36.48% (27) |
| $0.5 \leq \hat{p} < 1$ | 31.08% (23) |
| $\hat{p} = 1$ | 32.43% (24) |

According to Table 4, In individuals, the awareness of self-control on average is 66% and the average discount factor is 0.858. WI of people is 4265.3 on average which is as same as their commitment cost.

**Table 4.** Results obtained from the first and second stage of the test

| Variable | Min value | Max value | Average | Std. |
|---|---|---|---|---|
| $\hat{p}$ | 0.044 | 1 | 0.66 | 0.32 |
| $\delta$ | 0.767 | 0.995 | 0.858 | 0.29 |
| WI | 1000 | 10000 | 4265.3 | 3060.33 |

---

[11] Present bias

In our test we ignore $V_f$ because of complexity and considered it's zero then in the result for some of the people (about 48) who didn't want pay anything for commitment and flexibility value of $\hat{p}$ became zero. According to equations relation of $\hat{p}$ with $V_f$ is linear. It can be a future subject for present a method for analyses $V_f$ for each person and calculate with it.

In the following of this section, we examine the model on the data of Casari's (2009) research. In Table 5 we categorized individuals in the test. Half of the individuals were male and others were female. About 13% of individuals select SS and took $100 with 0-4 days delay ($\epsilon$), it shows that temptation of pure money is less than the food and sweets that were tested in the previous section. The costless commitment had more demand among individuals and it was about 57.3% but the demand for flexibility with cost was 12.1%, it's a small value and the opposite of commitment. The costly commitment was also 17.1% and is small respect costless type of it.

**Table 5.** Subjects respect to selection reward types.

| %Gender (N) | Reward type | %Subject (N) |
|---|---|---|
| 50% Female (41) | SS | 13.4% (11) |
| | LL with costly commitment | 17.1% (14) |
| 50% Male (41) | LL with costless commitment | 57.3% (47) |
| | LL with flexibility | 12.1% (10) |

In Table 6, results obtained from the first stage of the test, the values of D* individuals, which indicate the number of days those individuals are indifferent to receiving SS and LL, have a median of 14 for all subjects and FD* of them have a median of 42. The cost of commitments is selected as a fixed value of 20,000 rials, and the cost of flexibility is also fixed value of 20,000 rials.

**Table 6.** Results obtained from the first and second stage of the test

| Variable | Median |
|---|---|
| D* | 14 |
| FD* | 42 |
| Commitment's cost | 20000 |
| Flexibility's cost | 20000 |

According to Table 7, among those who had low self-control and were not aware of it, all of subjects have less than 50% aware then we saw choice reversal action between them, also a large percentage of people like before test showed *partially naive* behavior with low awareness. Small value of $\hat{p}$ show less awareness and observed 82 people that small value of $\hat{p}$ reversed their choice. The amount of PB in individuals is about 0.8 to 0.95 and preferred 0.88 in average, we calculated values of $\hat{p}$ in terms of 0.88 and 0.95.

**Table 7.** Results obtained from the first and second stage of the test

| Range $\hat{p}$ | %Subjects* (N) | %Subjects** (N) |
|---|---|---|
| $0 < \hat{p} < 0.5$ | 100% (82) | 100% (82) |
| $0.5 \leq \hat{p} < 1$ | - | - |
| $\hat{p} = 1$ | - | - |

\* Calculate p of subject with $\beta = 0.88$
\*\* Calculate p of subject with $\beta = 0.95$

According to Table 8, In individuals, the awareness of self-control on average is 16% with $\beta = 0.88$ and 41% with $\beta = 0.95$ and the average discount factor is 1.002 and 0.996. WI of people is 20,000 Rials which is as same as their commitment cost.

**Table 8.** Results obtained from the first and second stage of the test

| Variable | Average ($\beta = 0.88$) | Average ($\beta = 0.95$) |
|---|---|---|
| $\hat{p}$ | 0.16 | 0.42 |
| $\delta$ | 1.002 | 0.996 |
| WI | 20000 | 20000 |

# 7    Conclusion

In this study, the discount rate and the awareness of the self-control problem for each person are calculated by the hyperbolic discounting model through the experimental test. Subject to information collected from 38 participants, about one-third of them chose SS in all cases in the first stage and the rest of the people preferred LL. The awareness of the self-control problem of individuals varies from 0.16 to 1, of which only two people have a $\hat{p}$ equal to 1 (completely sophisticated) and more than half of them have a self-control problem of more than 0.5 and close to one, which indicates high self-awareness. Although most participants fall into this category, only 28% of them want to have a commitment device, and 40% prefer to have flexibility. About 31% of individuals were tempted and preferred sweets at that time. Individuals in this test chose LL for reward value utility 48% respect SS. Also, individuals who demand commitment devices increased their WI about 9% respect SS on average just with payment of commitment devices. In mapping, the model to the Casari's test (subjects that changed their decision) results showed a low level of awareness to the reverse decision. On average, people become indifferent to the choice of SS and LL in one month in both tests. The results obtained from the second stage of the test to confirm the hyperbolic discounting theory, indicate that people are indifferent to receiving SS and LL by adding one week to the time of receiving the awards but in Casari's test it was about 50 days. In both tests, $\beta$ was used as an average value but in Casari's test for checking the result of awareness with the maximum value, we used 0.95, and also again the value of $\hat{p}$ was under 0.5 then the performance of the model about predict choice reversal action of subjects was acceptable. Our observations also show that the average minimum discount rate of individuals is 1.027 and we couldn't show a significant linear relationship between awareness of self-control problem with discount rate and D*.

## Appendix 1

**Proof 1**: If utility function is defined as: $U_t = u_0 + \beta\delta u_1 + \beta\delta^2 u_2 + \cdots + \beta\delta^t u_t$ and A and B is defined as options we have:

$A: u_0 = 0, u_1 = U_{SS}, u_2 = 0$
$B: u_0 = 0, u_1 = 0, u_2 = U_{LL}$

$U_A^{(0)} = u_0 + \beta\delta u_1 + \beta\delta^2 u_2 = 0 + \beta\delta U_{SS} + 0$
$U_B^{(0)} = u_0 + \beta\delta u_1 + \beta\delta^2 u_2 = 0 + 0 + \beta\delta^2 U_{LL}$

$$U_A^{(1)} = u_1 + \beta\delta u_2 = U_{SS} + 0$$
$$U_B^{(1)} = u_1 + \beta\delta u_2 = 0 + \beta\delta U_{LL}$$

**Proof 2**: $U_A^{(0)} < U_B^{(0)}$ and $U_A^{(1)} > U_B^{(1)}$ then we can conclude that: $\beta\delta U_{SS} < \beta\delta^2 U_{LL}$ and $U_{SS} > \beta\delta U_{LL}$

Then $\frac{U_{SS}}{\delta} < U_{LL} < \frac{U_{SS}}{\delta\beta}$

**Proof 3**:
$$E\left(U_B^{(0)}{}_{flex}\right) = \hat{p}U_A^{(0)} + (1-\hat{p})U_B^{(0)}$$

$$E\left(U_B^{(0)}{}_c\right) - M = E\left(U_B^{(0)}{}_{flex}\right)$$

This is for maximum M:
$$\beta\delta^2 U_{LL} - V_f - M = \hat{p}U_A^{(0)} + (1-\hat{p})U_B^{(0)}$$

$$\hat{p} = \frac{M + V_f}{\beta\delta^2 U_{LL} - \beta\delta U_{SS}} \ .\blacksquare$$